\documentclass[twocolumn]{aastex61}
\usepackage[utf8]{inputenc}
\usepackage{amsmath, amsfonts, amssymb, graphics,graphicx, wrapfig, nicefrac}
\usepackage{epsfig}
\usepackage{epstopdf}
\usepackage{multirow}
\usepackage{graphicx}
\usepackage{rotating}
\usepackage[normalem]{ulem}

\def\13co{$^{13}$CO}
\def\hcop{HCO$^+$}
\def\h2{H$_2$}
\def\cm2{$\mathrm{cm^{-2}}$}
\newcommand{\hi}{H\,{\uppercase\expandafter{\romannumeral1}}}
\newcommand{\kms}{$\mathrm{km\, s^{-1}\, }$}

\submitjournal{ApJL}

\shorttitle{Revealing The CO X-factor In Dark Molecular Gas}
\shortauthors{Luo et al.}

\begin{document}
\begin{sloppypar}

\title{Revealing The CO X-factor In Dark Molecular Gas through Sensitive ALMA Absorption Observations}


\author{Gan Luo}
\affiliation{CAS Key Laboratory of FAST, National Astronomical Observatories, Chinese Academy of Sciences, Beijing 100101, China; luogan@nao.cas.cn, dili@nao.cas.cn, nytang@nao.cas.cn}
\affiliation{University of Chinese Academy of Sciences, Beijing 100049, China}

\author{Di Li}
\affiliation{CAS Key Laboratory of FAST, National Astronomical Observatories, Chinese Academy of Sciences, Beijing 100101, China; luogan@nao.cas.cn, dili@nao.cas.cn, nytang@nao.cas.cn}
\affiliation{University of Chinese Academy of Sciences, Beijing 100049, China}
\affiliation{NAOC-UKZN Computational Astrophysics Centre, University of KwaZulu-Natal, Durban 4000, South Africa}

\author{Ningyu Tang}
\affiliation{CAS Key Laboratory of FAST, National Astronomical Observatories, Chinese Academy of Sciences, Beijing 100101, China; luogan@nao.cas.cn, dili@nao.cas.cn, nytang@nao.cas.cn}

\author{J. R. Dawson}
\affiliation{Department of Physics and Astronomy and MQ Research Centre in Astronomy, Astrophysics and Astrophotonics, Macquarie University, NSW 2109, Australia}

\author{John M. Dickey}
\affiliation{University of Tasmania, School of Maths and Physics, Hobart, TAS 7001, Australia}

\author{L. Bronfman}
\affiliation{Astronomy Department, Universidad de Chile, Casilla 36-D, Santiago, Chile}

\author{Sheng-Li Qin}
\affiliation{Department of Astronomy, Yunnan University, and Key Laboratory of Astroparticle Physics of Yunnan Province, Kunming, 650091, China}

\author{Steven J. Gibson}
\affiliation{Western Kentucky University, Dept. of Physics and Astronomy, 1906 College Heights Boulevard, Bowling Green, KY 42101, USA}

\author{Richard Plambeck}
\affiliation{Radio Astronomy Laboratory, University of California, Berkeley, CA 94720}

\author{Ricardo Finger}
\affiliation{Astronomy Department, Universidad de Chile, Casilla 36-D, Santiago, Chile}

\author{Anne Green}
\affiliation{Sydney Institute for Astronomy (SIfA), School of Physics, University of Sydney, NSW 2006, Australia}

\author{Diego Mardones}
\affiliation{Departamento de Astronom{\'i}a, Universidad de Chile, Casilla 36, Santiago de Chile, Chile}

\author{Bon-Chul Koo}
\affiliation{Department of Physics and Astronomy, Seoul National University Seoul 151-747, Korea}

\author{Nadia Lo}
\affiliation{Astronomy Department, Universidad de Chile, Casilla 36-D, Santiago, Chile}

\begin{abstract}
Carbon-bearing molecules, particularly CO, have been widely used as tracers of molecular gas in the interstellar medium (ISM). In this work, we aim to study the properties of molecules in diffuse, cold environments, where CO tends to be under-abundant and/or sub-thermally excited. We performed one of the most sensitive (down to $\mathrm{\tau_{rms}^{CO} \sim 0.002}$ and $\mathrm{\tau_{rms}^{HCO^+} \sim 0.0008}$) sub-millimeter molecular absorption line observations towards 13 continuum sources with the ALMA.
CO absorption was detected in diffuse ISM down to $\mathrm{A_v< 0.32\,mag}$ and \hcop was down to  $\mathrm{A_v < 0.2\,mag}$, where atomic gas and dark molecular gas (DMG) starts to dominate. Multiple transitions measured in absorption toward 3C454.3 allow for a direct determination of excitation temperatures $\mathrm{T_{ex}}$ of 4.1\,K and 2.7\,K, for CO and for \hcop, respectively, which are close to the cosmic microwave background (CMB) and provide explanation for their being undercounted in emission surveys. A stronger linear correlation was found between $\mathrm{N_{HCO^+}}$ and $\mathrm{N_{H_2}}$ (Pearson correlation coefficient P $\sim$ 0.93) than that of $\mathrm{N_{CO}}$ and $\mathrm{N_{H_2}}$ (P $\sim$ 0.33), suggesting \hcop\ being a better tracer of H$_2$ than CO in diffuse gas.
The derived CO-to-\h2 conversion factor (the CO X-factor) of (14 $\pm$ 3) $\times$ 10$^{20}$ cm$^{-2}$ (K \kms)$^{-1}$ is approximately 6 times larger than the average value found in the Milky Way.
\end{abstract}


\section{Introduction}\label{sec:intro}
Hydrogen is the most abundant element in the universe and plays a crucial role in the chemical evolution of cosmic materials. In contrast to the 21\,cm hyperfine transition of atomic hydrogen, molecular hydrogen has no easily accessible transitions (due to a lack of permanent dipole) in the cold (T$\sim$15\,K) ISM, where H$_2$ is the dominant component of gases.
CO is the de-facto stand-in for H$_2$ in both the Milky Way and external galaxies, ever since its discovery \citep{Wilson1970}. Total molecular gas is often estimated through an empirical CO-to-\h2 conversion factor (hereafter, the CO X-factor), namely, $\mathrm{N_{H_2}=X_{CO}W_{CO}}$, a typical mean value of which  is 2 $\times$ 10$^{20}$ cm$^{-2}$ (K \kms)$^{-1}$ for the Milky Way \citep{Bolatto2013}.

However, CO is not always a reliable tracer of H$_2$. A significant fraction of interstellar molecular gas is not traced by CO emission surveys (e.g.\  \citealt{Dame2001}) and is often referred to as dark molecular gas (DMG). DMG has been revealed by gamma-ray observations from the Energetic Gamma Ray Experiment Telescope (EGRET, \citealt{Grenier2005}) and Fermi \citep{Remy2017}, dust observations with Planck \citep{Planck2011}, and C$^+$ observations from Galactic observations of TeraHertz C+ (GOTC+) survey \citep{Pineda2013,Langer2014}. CO emission fails to trace DMG in two critical aspects, abundance and excitation.
The self-shielding of H$_2$ can be achieved at much lower extinction than that for CO. In regions with intermediate extinctions ($\mathrm{A_v \sim 0.2--2\,mag}$ see \citealt{Li2018}), ultraviolet dissociation significantly reduces the CO abundance from its cannonical value of [CO]/[H$_2$] = $10^{-4}$ \citep{van1988,Wolfire2010,Draine2011}.
In such regions, the gas density can be lower than the critical density, making transitions such as CO (1--0) sub-thermally excited, further hampering the usability of CO emission as a H$_2$ tracer.

The ``missing" CO emission in diffuse and translucent clouds may lead to a higher CO X-factor than in dense molecular gas.
Observational evidence from dust extinction and CO emission has revealed a larger CO X-factor in diffuse regions ($\mathrm{A_v < 3\,mag}$) than in dense gas \citep{Pineda2010} in the Taurus molecular cloud. A similar trend has also been found by \citet{Remy2017} toward six nearby anti-centre clouds.
However, other studies have reported a mean value of the CO X-factor that is the same in diffuse and dense molecular clouds \citep{Liszt2010,Liszt2012}, although there are considerable variations between individual sightlines.

Given the high temperatures of background sources, quasar absorption can be a much more sensitive and less biased probe of cold molecular gas, particularly, the DMG in the intermediate extinction regions. Absorption toward continuum sources has been commonly used to quantify the physical properties of the Galactic interstellar medium \citep[e.g.][]{Heiles2003a,Dickey2013,Li2018,Nguyen2018,Riquelme2018}.
In such absorption programs, it is possible to measure the column density accurately even when a transition is sub-thermally excited, to detect diffuse gas with lower column density, and to directly measure optical depth. Furthermore, absorption from multiple transitions facilitate an unambiguous determination of excitation temperature and column density simultaneously.

In addition to CO, \hcop\ is a key reactant of CO formation, and has been shown to trace H$_2$ well \citep{Liszt1998}.
The emission of \hcop\ is rare and weak in diffuse clouds \citep[tens of mK in brightness temperature,][]{Lucas1996}, and excitation temperatures derived from the two lowest transitions are comparable to the CMB \citep[2.7--3.0\,K,][]{Godard2010}, making absorption toward strong continuum sources a preferred probe of \hcop as well.

We have taken deep absorption observations of CO and \hcop\ toward 13 millimeter-wavelength point source calibrators using the ALMA. 
We present our observations and archival data in Section \ref{sec:observation}, and analyze the absorption profiles of the observed targets in Section \ref{sec:abs profile}. We derive the abundance of CO and \hcop and discuss the advantage of \hcop as a tracer of molecular gas in Section \ref{sec:hcop}. Analysis of the CO X-factor is presented in Section \ref{sec:x-factor}. Our main results and conclusions are presented in Section \ref{sec:conclusion}.

\section{Observations and data}\label{sec:observation}

We have selected 13 calibration sources (3C454.3, 3C120, B0420-014, B0607-157, B0627-199, B1730-130, B1741-038, B1742-078, B1933-400, B1954-388, B2223-052, B2227-088, and B2243-123) used by worldwide mm-wave arrays. In addition, none of these sightlines has CO emission in the CO survey in \citealt{Dame2001}. The sightlines of these sources lie in DMG \citep[$\mathrm{A_v: 0.37\,mag \sim 2.5\,mag}$]{Planck2011} or DMG-threshold (as defined in \citealt{Li2018}) regions, enabling us to explore the physical properties in DMG sightlines. 

\subsection{Molecular lines}\label{sec:molecular lines}
We observed CO (1--0) and \hcop\ (1--0) spectra toward 13 strong continuum sources with ALMA during 2015. We also observed two higher energy transitions, CO (2--1) and \hcop\ (3--2), toward 3C454.3 to constrain the physical properties (see Section \ref{sec:analysis}). The frequency resolution of the CO (1--0) and \hcop\ (1--0) lines is 61.035\,KHz, which corresponds to velocity resolutions of 0.21 \kms at 89\,GHz and 0.18 \kms at 113\,GHz. The frequency resolution of the CO (2--1) and \hcop\ (3--2) lines is 122.070\,KHz, which corresponds to velocity resolutions of 0.159 \kms at 230\,GHz and 0.137 \kms at 267\,GHz.
The continuum data covers a 1.8\,GHz bandwidth with a frequency resolution of 31 MHz. The calibration of the raw data was performed using CASA (Common Astronomy Software Applications) \citep{MuMullin2007}. Both calibrated continuum and spectral images were deconvolved with Briggs weighting using the Clean algorithm.

\subsection{Archival reddening data}\label{sec:dust reddening}
We use the E(B-V) map of \citet{Green2018,Green2019} to 
estimate total gas column density. The E(B-V) map is derived by combining high-quality stellar photometry of 800 million stars from Pan-STARRS 1, 200 million stars from 2MASS, and parallaxes of 500 million stars from Gaia \citep{Green2019}. It covers three-quarters of the whole sky at a spatial resolution of 3.4$'$ to 13.7$'$.

\subsection{Archival \hi\ data}\label{sec:HI}
We adopted the corrected \hi\ column density derived from ON-OFF observations in the millennium survey for 3C454.3 and 3C120 \citep{Heiles2003a,Nguyen2018}. These \hi\ spectra have a velocity resolution of 0.161 \kms and an RMS noise of 56\,mK per channel. \hi\ data for other sources are taken from the Parkes Galactic All Sky Survey (GASS) \citep{McClure2009,Kalberla2010,Kalberla2015}, in which the column density of \hi\ is calculated by integration over a velocity range of -400 $<$ V $<$ 400 \kms under the assumption of $\tau \ll 1$, and is therefore a lower limit. The GASS data have an angular resolution of 16$'$, a spectral resolution of 0.82 \kms and an RMS noise of 57\,mK per channel. In addition, \hi\ data for two bandpass calibrators (2148+0657 and 2232+1143; see Section \ref{sec:selfcal}) are taken from the Effelsberg-Bonn \hi\ Survey (EBHIS), which has an angular resolution of $\sim 11'$, a spectral resolution of 1.29 \kms and an RMS noise of 90\,mK per channel \citep{Winkel2016}.

\section{Absorption Profiles} \label{sec:abs profile}
Absorption profiles can provide a direct measurement of optical depth, provided that the background source is sufficiently bright. Optical depth can tell us the column density along a sightline, with the proper assumption of the excitation conditions. In this section, we describe the extraction of absorption spectra and the fitting process.

\subsection{Detections and self-calibration}\label{sec:selfcal}
Since our observing targets are strong millimeter-wavelength continuum point sources, we can perform calibration using the sources themselves (self-calibration). 
Eliminating separate calibrator observations would have cut our observing time by more than half. However, ALMA Phase 2 observations did not support this mode. Thus, we performed our calibration using the standard ALMA pipeline, which uses separate bandpass, phase, and flux calibrators.

\begin{figure*}
\gridline{\fig{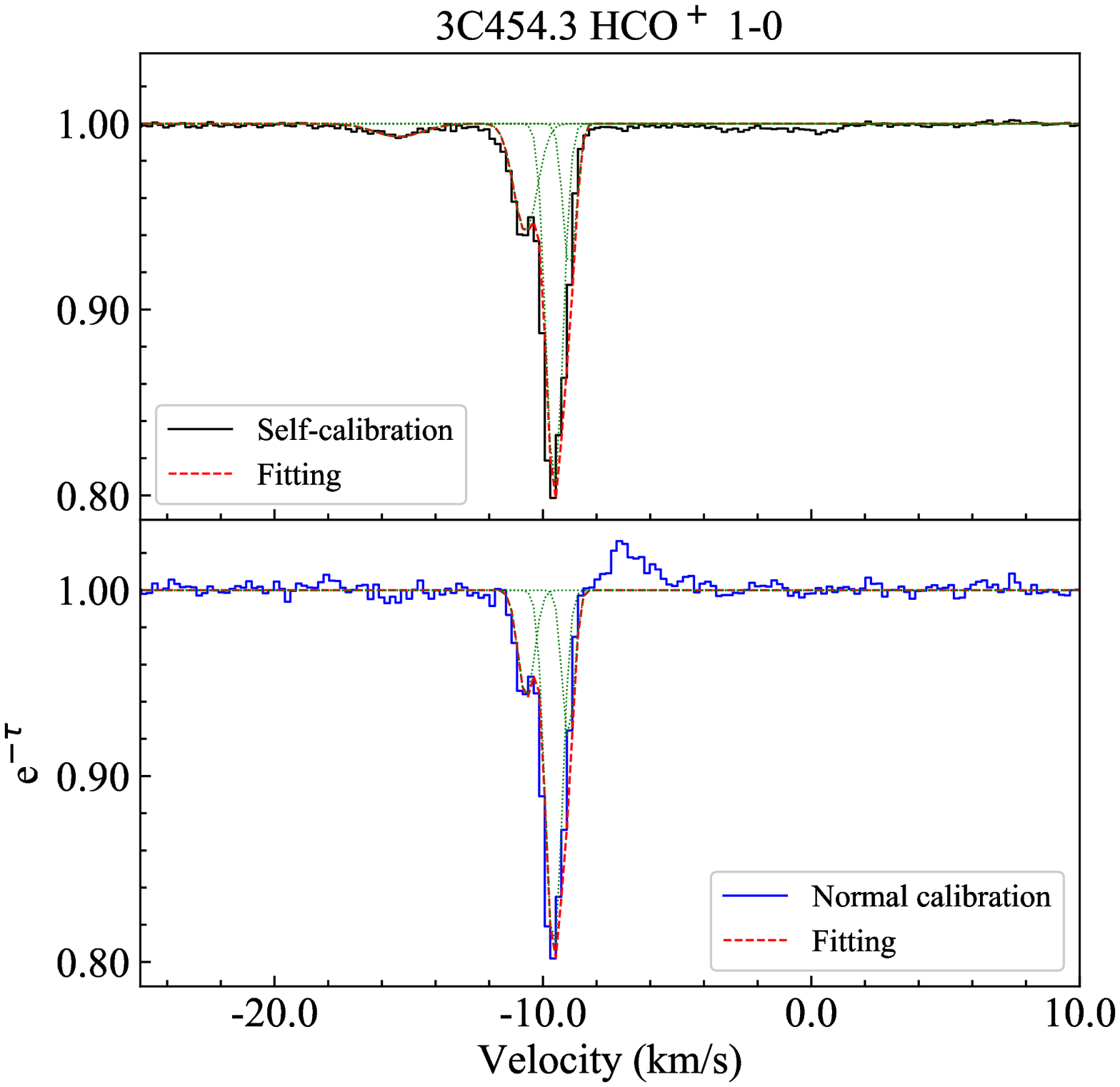}{0.5\textwidth}{(a)}
          \fig{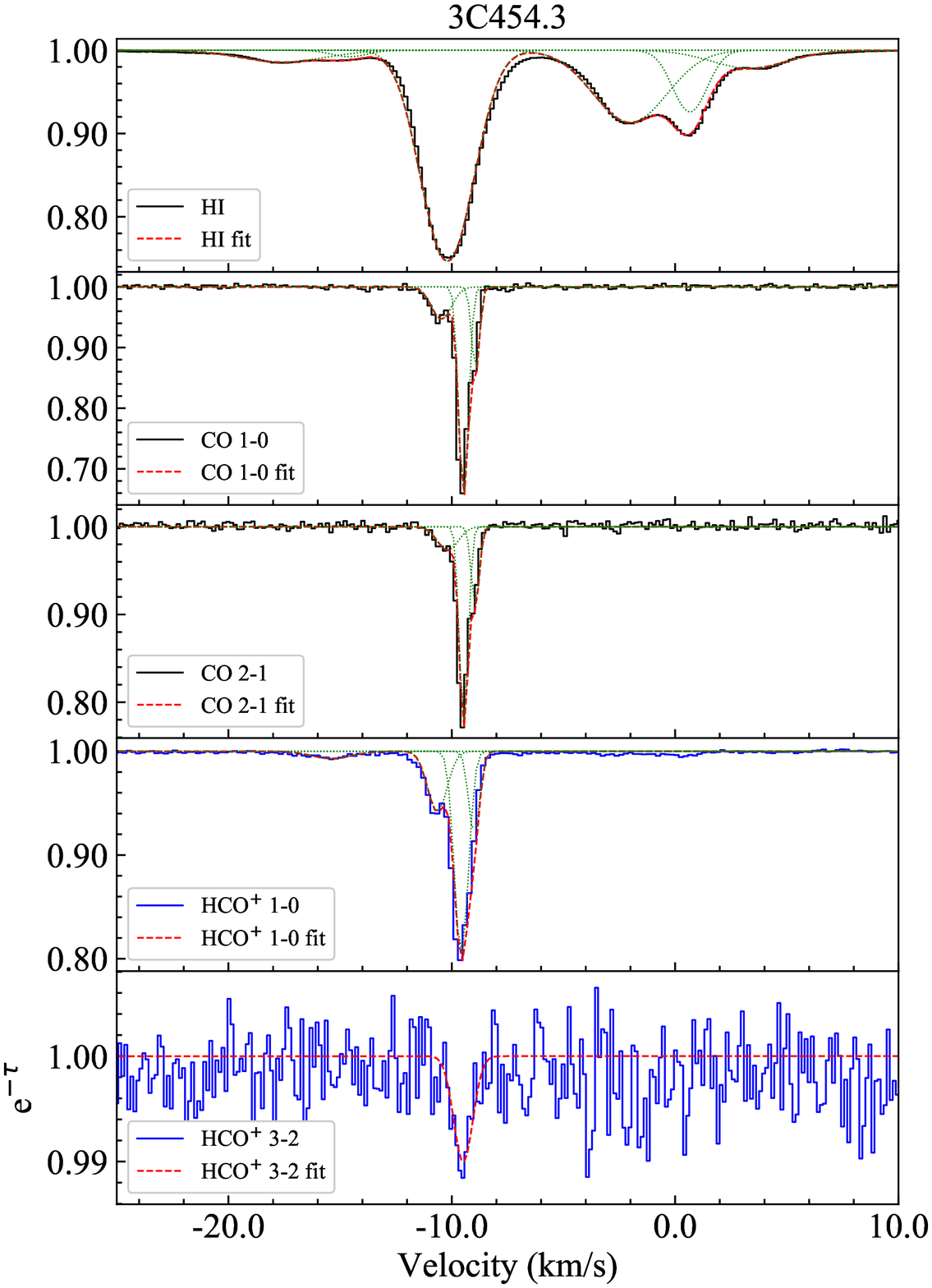}{0.5\textwidth}{(b)}
          }
\caption{(a): Comparison of absorption spectra obtained through self- (black curve) and normal (blue curve) calibration. (b): The spectra of \hi, CO and \hcop\ (solid line) towards 3C454.3. The red dashed lines show the Gaussian fit profiles for each line, and the green dotted lines show each of the Gaussian components. The x-axis and y-axis denote the velocity range in \kms and the normalized line intensity, respectively. \label{fig:lines}}
\end{figure*}

Both CO (1--0) and \hcop\ (1--0) absorption lines were detected toward 6 of 13 sources: 3C454.3, 3C120, 0607-157, 1730-130, 1741-038, and 1742-078. 
The two higher energy transitions -- CO (2--1) and \hcop\ (3--2) -- observed toward 3C454.3 were also detected.
In our \hcop\ (1--0) absorption profiles, we found two emission components toward 3C120 and 3C454.3 (see Figure \ref{fig:lines} (a) as an example). High brightness temperatures of 108\,K (for 3C120) and 537\,K (for 3C454.3), make it hard to believe that the emission could be excited by high-temperature gas, and suggest contamination from a bandpass calibrator. Thus, we performed self-calibration on 3C120 and 3C454.3, as well as their calibrators.
As suspected, \hcop\ (1--0) absorption was detected towards two bandpass calibrators (2232+1143 and 2148+0657), confirming that the ``emission'' components above are indeed contamination. 
The self-calibrated and normally-calibrated spectra for 3C454.3 can be seen in Figure \ref{fig:lines} (a).
Data reduction toward 3C454.3, 3C120, (and also 2148+0657 and 2232+1143, which were added to our sample) was therefore conducted by self-calibration. 
Calibration toward all the other sources was conducted by normal calibration.

\subsection{Gaussian fitting of absorption profiles}\label{sec:gaussian fit}
According to the radiative transfer equation:
\begin{equation}
T^*_{R} = J_{\upsilon}(T_{bg})e^{-\tau_\upsilon} + J_{\upsilon}(T_{ex})(1-e^{-\tau_\upsilon}) ,
\label{eq:tmb}
\end{equation}
where $\mathrm{T^*_{R}}$ is the main beam brightness temperature, $\mathrm{\tau_\upsilon}$ is the optical depth, and $\mathrm{J_{\upsilon}(T) = (h\upsilon/k)/(exp(h\upsilon/kT)-1)}$.
We can then express $\mathrm{e^{-\tau_\upsilon}}$ as
\begin{equation}
e^{-\tau_\upsilon} = \frac{T^*_{R}-J_{\upsilon}(T_{ex})}{J_{\upsilon}(T_{bg})-J_{\upsilon}(T_{ex})} .
\label{eq:tau1}
\end{equation}
When the background continuum source is strong enough (e.g., 3C454.3 with $\mathrm{J_{\upsilon}(T_{bg}) \sim 537\,K}$), term $\mathrm{J_{\upsilon}(T_{bg}) \gg J_{\upsilon}(T_{ex})(1-e^{-\tau_\upsilon})}$, and equation \ref{eq:tau1} can be rewritten as
\begin{equation}
e^{-\tau_\upsilon} \simeq \frac{T^*_{R}}{J_{\upsilon}(T_{bg})}.
\label{eq:tau2}
\end{equation}

The $\mathrm{\tau_\upsilon}$ of each velocity component was derived by fitting the $\mathrm{e^{-\tau_\upsilon}}$ profiles with multiple Gaussian components.
Equation \ref{eq:tau2} is only used for obtaining the $\mathrm{\tau_\upsilon}$ value of 3C454.3. 
The other sources observed in this study have background brightness temperatures between 20\,K to 176\,K, meaning that we must revert to equation \ref{eq:tau1} to compute $\mathrm{e^{-\tau_\upsilon}}$ profiles. For these, we adopt representative values of $\mathrm{T_{ex}}$ computed for 3C454 (4.1\,K for CO and 2.7\,K for \hcop\ in Section \ref{sec:analysis}). The $\mathrm{e^{-\tau_\upsilon}}$ profiles of 3C454.3 are shown in Figure \ref{fig:lines} (b).

There are twelve CO (1--0) components and twenty-three \hcop\ (1--0) components in our sample. Each CO (1--0) velocity component has a corresponding \hcop\ (1--0) component. Those sightlines with CO detection suggesting that CO has been formed in diffuse gas down to $\mathrm{A_v}$ value of 0.32\,mag (converting from E(B-V) in Table \ref{tab:abundance} using $\mathrm{A_v/E(B-V) = 3.1}$). Eleven \hcop\ components have no CO detections, suggesting that these components lie in diffuse molecular gas where CO cannot form and that they may lie at the edge of the molecular cloud. Specifically, there are two sightlines without CO detections but which are rich in \hcop, indicating that HCO$^+$ exists down to $\mathrm{A_v}$ value of 0.19\,mag. 

The velocity of these absorption profiles varies from $\sim$ -10--10 \kms, indicating that the cloud components along the sightline are nearby clouds. The optical depths of CO(1--0) and \hcop(1--0) are in the range of $\sim$0.04--1.55 and $\sim$0.007--0.98, respectively. The linewidths of \hcop\ (0.56--3.96 \kms) are sightly wider than CO (0.21--3.79 \kms) by a factor of 1.05, which is consistent with the results of \citealt{Liszt1998}.

\section{Radiative transfer analysis}\label{sec:analysis}
Multiple absorption lines from CO and \hcop\ allow us to constrain the physical properties along a sightline. Given the specific input parameters (kinetic temperature, $\mathrm{T_{kin}}$, volume density of collision partners, $n$ (e.g. p-\h2, o-\h2, e$^-$), column density of specific molecule $N_\mathrm{mol}$, and full width at half maximum (FWHM) of a specific line, $\mathrm{\delta \upsilon}$), we are able to model the observed spectra with RADEX \citep{Van2007}.

In this work, we considered p-\h2 and o-\h2 ([o-\h2]/[p-\h2] = 10$^{-3}$; \citealt{Bourlot1991,Dislaire2012}) as the collision partners of CO, and \h2 and electrons \citep[$\mathrm{\left [ e^-/H_2 \right ] = 10^{-4}}$;][]{Bhattacharyya1981,Lucas1996,Liszt2012} as the collision partners of \hcop.
Adopting a kinetic temperature of 100\,K for the diffuse ISM \citep{Goldsmith2013,Gerin2015}, we vary the free parameters $\mathrm{n_{H_2}}$, $\mathrm{N_{CO}}$, and $\mathrm{N_{HCO^+}}$ to find the optimum solutions by minimizing $\chi^2$ for the four absorption spectra of CO and \hcop\ toward 3C454.3. Here $\chi^2$ is defined as
\begin{equation}
\chi^2=\frac{1}{n}\sum_{i=1}^{n} \frac{\left ( \tau^i_{model}-\tau^i_{obs} \right )^2}{{\sigma^i_{obs}}^2} ,
\label{eq:kai}
\end{equation}
where $\mathrm{\tau^i_{model}}$ is the output optical depth generated by RADEX, $\mathrm{\tau^i_{obs}}$ is the observed optical depth of the four spectra, and $\mathrm{\sigma^i_{obs}}$ is the uncertainty of the observed optical depth for each transition.

For the other sources, we have only one transition for CO and one for \hcop. The same kinetic temperature (100\,K) and the $\mathrm{n_{H_2}}$ obtained for 3C454.3 (88 $\mathrm{cm^{-3}}$) were then used as the initial assumptions when modeling the optical depth toward the other sources.

Analyzing the fitting results, we obtain the following results:
\begin{itemize}
    \item The best fit values for $\mathrm{n_{H_2}}$, $\mathrm{N_{CO}}$, and $\mathrm{N_{HCO^+}}$ toward 3C454.3 are 87 $\mathrm{cm^{-3}}$, 3.6$\times 10^{14}$ \cm2, and 1.8$\times 10^{11}$ \cm2, respectively. The derived excitation temperature values are 4.1\,K for CO and 2.7\,K for \hcop. Our results are consistent with \citealt{Goldsmith2013} ($\mathrm{T_{ex}(CO) \sim 2.7-13.6\,K}$) and \citealt{Godard2010} ($\mathrm{T_{ex}(CO) \sim 2.7-3.0\,K}$).
    \item By varying the assumed $\mathrm{T_{kin}}$ value from 10 to 300\,K, we found that the fractional difference are within 1$\%$ for $\mathrm{T_{ex} (CO)}$ and $\mathrm{N_{CO}}$, and 0.1$\%$ for $\mathrm{T_{ex} (HCO^+)}$ and $\mathrm{N_{HCO^+}}$.
    \item The column densities of CO and \hcop\ components in our sample are in the range of (0.1--19) $\times 10^{14}$ cm$^{-2}$ and (0.07--8.8) $\times 10^{11}$ cm$^{-2}$, respectively. Three $\sigma$ upper limits in CO column density are given for those components without CO (1--0) detections (red points with arrows in Figure \ref{fig:CO_HCO+}). As shown in Figure \ref{fig:CO_HCO+} (a), the column densities of CO and \hcop\ show a positive correlation. The ratio of $\mathrm{N(HCO^+)/N(CO)}$ decreases by an order of magnitude as the CO column density goes from lower to higher, as shown in Figure \ref{fig:CO_HCO+} (b). Similar results have also been reported by \citealt{Turner1995}, who found that the $\mathrm{N(HCO^+)/N(CO)]}$ ratio is 10 times higher toward regions with low extinction ($\mathrm{A_v < 1\,mag}$) than that of higher extinction ($\mathrm{A_v > 1.5\,mag}$). Our samples complement the data points of \citet{Lucas1996}, extending the relation to a much lower column density (down to $\mathrm{7 \times 10^9 cm^{-2}}$ for \hcop).
    \begin{figure*}
    \gridline{\fig{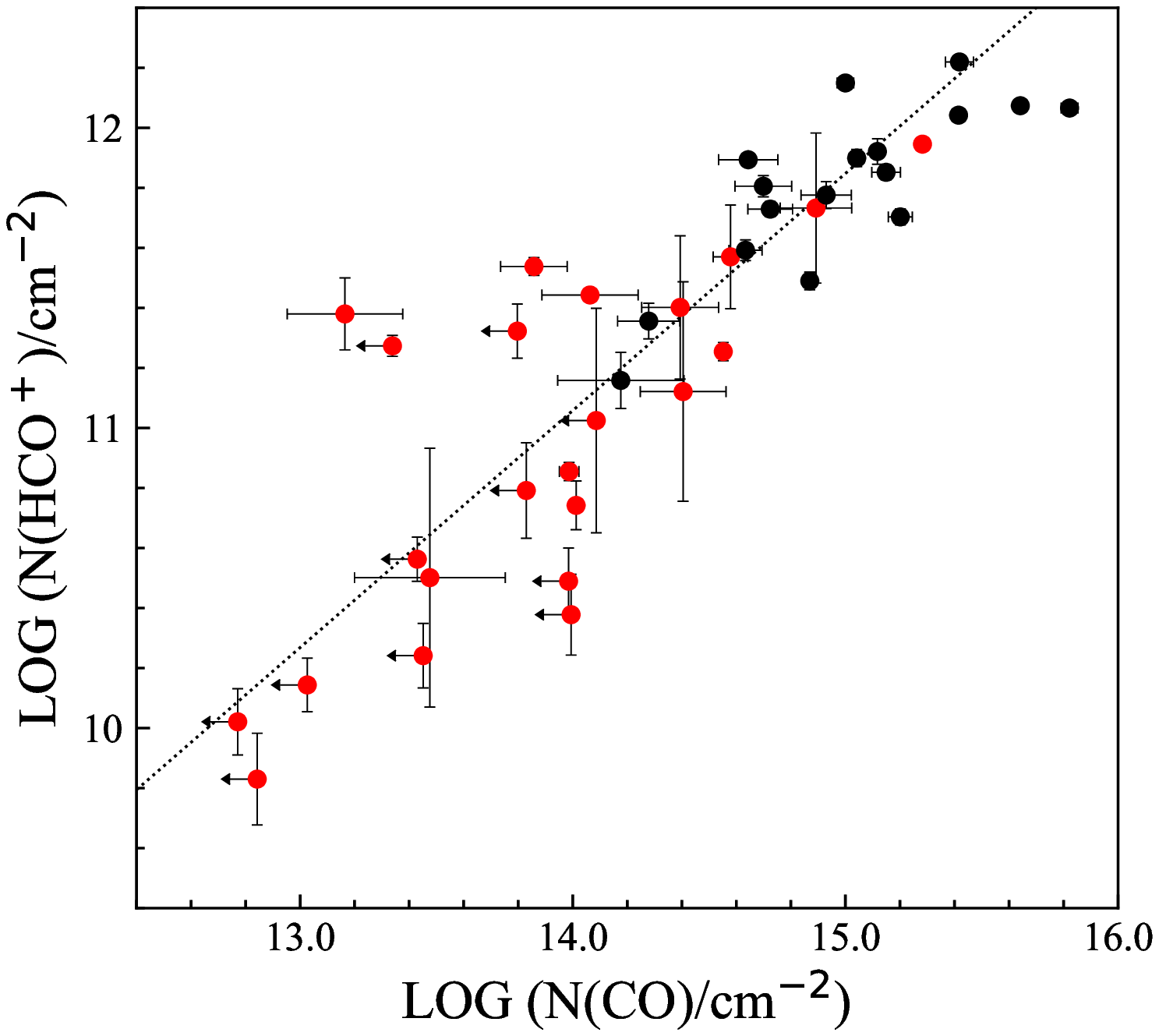}{0.5\textwidth}{(a)}
            \fig{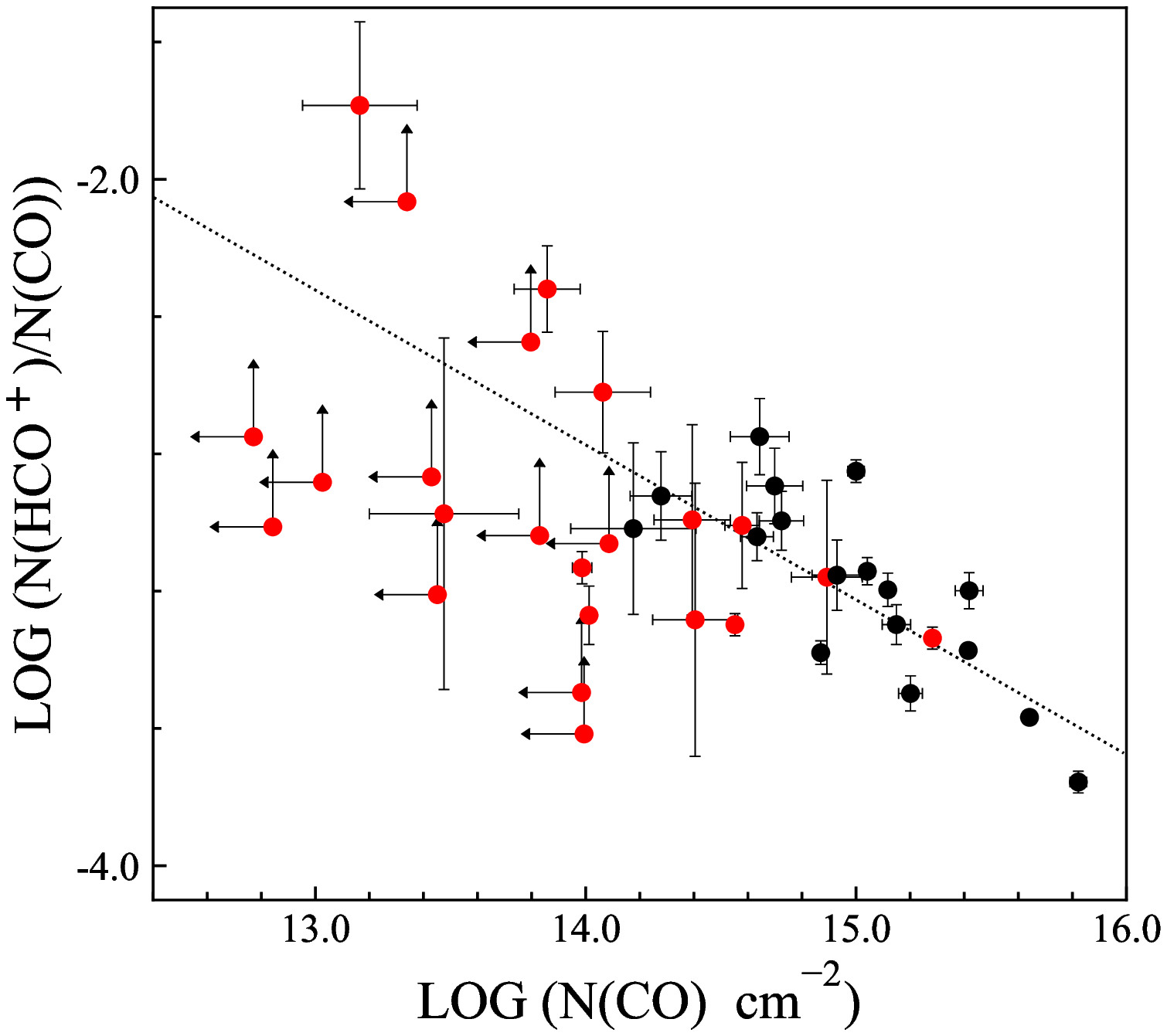}{0.5\textwidth}{(b)}
          }
    \caption{(a): \hcop\ column density vs. CO column density in a log-log scale; (b): $\mathrm{ N(HCO^+) / N(CO)}$ ratio vs. CO column density in a log-log scale. The red dots are from our work and black dots are from \citet[CO column density in Table 5]{Liszt1998} and \citet[\hcop\ column density in Table 2]{Lucas1996}. The black dotted lines show the fit slope to the data points from all components (upper limits are not included).}
    \label{fig:CO_HCO+}
    \end{figure*}
\end{itemize}

\section{\hcop\ as a molecular gas tracer}\label{sec:hcop}

In this section, we derive the relative abundance of CO and \hcop\ with respect to H$_2$ by $\mathrm{N_{H_2} = \frac{N_H - N_{\hi\ }}{2}}$. Here, $\mathrm{N_H}$ is estimated from $\mathrm{E(B-V)}$ (shown in Table \ref{tab:abundance}), as described below. $\mathrm{N_{\hi}}$ is obtained as described in Section \ref{sec:observation}, and for the majority of sightlines represents a lower limit under the optically thin assumption.

Interstellar reddening is caused by dust absorption and scattering.
Observations from Ly$\alpha$ absorption toward 100 stars showed a linear relation between total gas column density and interstellar reddening: $\mathrm{N_H /E(B-V) = 5.8 \times 10^{21} cm^{-2} mag^{-1}}$ \citep{Bohlin1978}. The ratio has been widely used to convert reddening to total gas column density. However, different $\mathrm{N_H (cm^{-2})/E(B-V)}$ ratios have been found in different environments by recent observations. \citet{Planck2014} compared reddening from Sloan Digital Sky Survey (SDSS) measurements of quasars with total gas column density in the diffuse ISM, and found a value $\sim$1.2 times larger than in \citet{Bohlin1978}. \citet{Liszt2014a,Liszt2014b} compared $\mathrm{N_{\hi}}$ and $\mathrm{E(B-V)}$, resulting in $\mathrm{N_H (cm^{-2})/E(B-V)=8.3 \times 10^{21} cm^{-2} mag^{-1}}$ for $\mathrm{\left | b \right | \geq 20^{\circ}}$ and $\mathrm{0.015 \leq E(B-V) \leq 0.075}$. Similar results have also been found by \citet{Lenz2017} in low $\mathrm{N_{\hi}}$ ($\mathrm{< 4 \times 10^{20} cm^{-2}}$) regions, and by \citet{Nguyen2018} along purely atomic sightlines ($\mathrm{8.8 \times 10^{21}}$ and $\mathrm{(9.4 \pm 1.6) \times 10^{21} cm^{-2} mag^{-1}}$, respectively). $\gamma$-ray observations from Fermi show the ratio decreases from atomic clouds to dense molecular clouds by a factor of $\sim$0.6 \citep{Remy2018}. In our case, the $\mathrm{E(B-V)}$ values along sightlines are in the range of 0.06 to 0.67\,mag ($\mathrm{A_v: 0.2 \sim 2\,mag}$), including both diffuse, atomic-dominated and translucent molecular gas environments. We still adopt the value $\mathrm{N_H (cm^{-2})/E(B-V)}$ derived by \citealt{Bohlin1978} for sightlines with $\mathrm{E(B-V)>0.5\,mag}$ (1730-130, 1741-038, and 1742-078). However, this ratio will lead to $\mathrm{N_H<N_{\hi}}$ towards atomic dominated sightlines (3C454.3, 3C120, 2148+0657, and 2232+1143). Thus, we adopt the value of $\mathrm{(9.4 \pm 1.6) \times 10^{21} cm^{-2} mag^{-1}}$ for sightlines with $\mathrm{E(B-V)<0.5\,mag}$ (Table \ref{tab:abundance}). 

\begin{figure}
\includegraphics[width=1.0\linewidth]{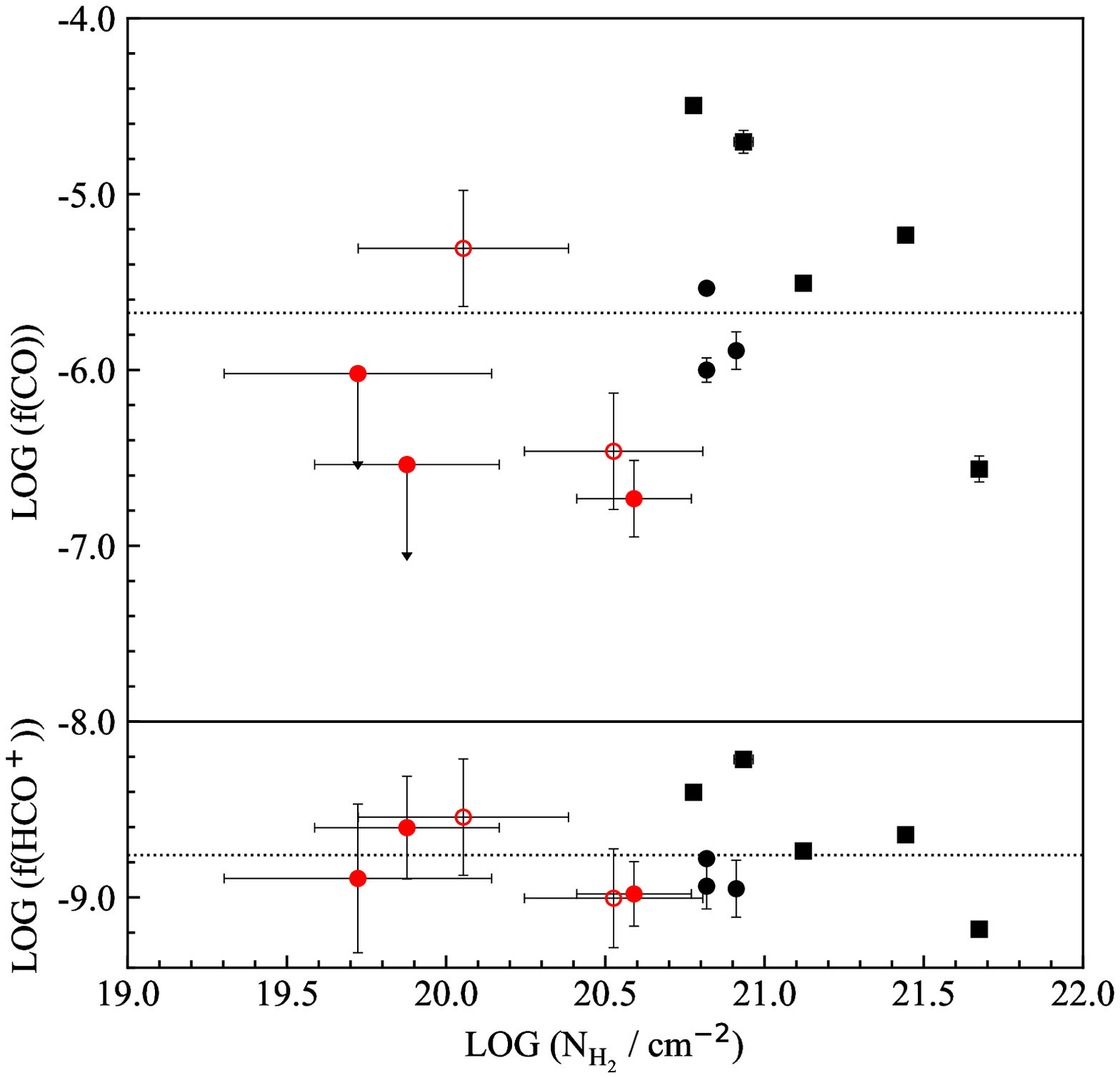}
\caption{The relation between the abundance of CO, \hcop\ and $\mathrm{N_{H_2}}$ (circles from our work, squares from \citealt{Lucas1996,Liszt1998}). Red and black points denote sightlines using $\mathrm{N_H /E(B-V) }$ conversion factor of \citealt{Nguyen2018} and \citealt{Bohlin1978}, respectively. Empty red circles denote sightlines with \hi\ opacity correction. The black dotted lines show the mean values of molecular abundance (points with upper limits are not included).}
\label{fig:abundance}
\end{figure}

The relation between the abundance of CO and \hcop\ and H$_2$ column density is shown in Figure \ref{fig:abundance} (datapoints in \citealt{Lucas1996,Liszt1998} are included). The abundance of CO ranges from (0.2$\pm$0.1) $\times$ 10$^{-6}$ to (5$\pm$4) $\times$ 10$^{-6}$, which is consistent with previous work from far-ultraviolet spectra of the diffuse ISM (mean value of $\sim$ 3 $\times$ 10$^{-6}$ in \citealt{Burgh2007}). This value is much lower than typical values in dense molecular clouds \citep[e.g., $10^{-5} \sim 10^{-4}$ in Taurus,][]{Goldsmith2008}.
The abundance of \hcop\ ranges from (1.0$\pm$0.4) $\times$ 10$^{-9}$ to (3$\pm$2) $\times$ 10$^{-9}$. The mean values of molecular abundance are $\mathrm{(2.1 \pm 1.1) \times 10^{-6}}$ for CO and $\mathrm{(1.7 \pm 0.3) \times 10^{-9}}$ for \hcop (upper limits are not included). The Pearson correlation coefficient values between the molecular column density and $\mathrm{N_{H_2}}$ are 0.33 for CO and 0.93 for \hcop, indicating that \hcop\ is a better tracer of molecular gas than CO.

\begin{deluxetable*}{cccccccccc}
\tablecaption{The E(B-V), $\mathrm{N_H}$, $\mathrm{N_{\hi\ }}$, $\mathrm{N_{H_2}}$, atomic fraction, abundance of CO and \hcop\, and rms towards 8 sources \label{tab:abundance}}
\tablewidth{700pt}
\tabletypesize{\scriptsize}
\tablehead{
\colhead{Source}& \colhead{E(B-V)} & \colhead{$\mathrm{N_H}$} & \colhead{$\mathrm{N_{\hi\ }}$} & \colhead{$\mathrm{N_{H_2}/ 10^{20} cm^{-2}}$} & \colhead{$\mathrm{f_{atomic}}$} & \colhead{$\mathrm{f_{CO}}$} & \colhead{$\mathrm{f_{HCO^+}}$} & \multicolumn{2}{c}{rms} \\
\colhead{}& \colhead{mag} & \colhead{$\mathrm{10^{21} cm^{-2}}$} & \colhead{$\mathrm{10^{20} cm^{-2}}$} & \colhead{$\mathrm{10^{20} cm^{-2}}$} & & \colhead{$10^{-6}$} & \colhead{$10^{-9}$} & \colhead{$\mathrm{\tau_{CO}}$} & \colhead{$\mathrm{\tau_{HCO^+}}$}
}
\startdata
3C454.3&0.106$\pm$0.002&1.0$\pm$0.2\tablenotemark{a}& 7.7$\pm$0.1\tablenotemark{c} & 1.1 $\pm$ 0.9 & 0.7$\pm$0.1 & 5 $\pm$ 4 & 3$\pm$2 & 0.003 & 0.001\\
3C120&0.265$\pm$0.006&2.5$\pm$0.4\tablenotemark{a}& 18.2$\pm$0.1\tablenotemark{c} & 3$\pm$2 & 0.7$\pm$0.1 & 0.5$\pm$0.3 & 1.0$\pm$0.6 & 0.02 & 0.004\\
0607-157&0.203$\pm$0.004&1.9$\pm$0.3\tablenotemark{a}& 11.3$\pm$0.2 & 4 $\pm$ 2 & 0.6$\pm$0.1 & 0.2$\pm$0.1 & 1.0$\pm$0.4 & 0.01 & 0.009\\
1730-130&0.513$\pm$0.009&2.98$\pm$0.05\tablenotemark{b}& 16.6$\pm$0.3 & 6.6 $\pm$ 0.3 & 0.56$\pm$0.01 & 2.9 $\pm$ 0.2 & 1.7$\pm$0.1 & 0.008 & 0.01\\
1741-038&0.530$\pm$0.008&3.07$\pm$0.05\tablenotemark{b}& 17.6$\pm$0.4 & 6.6 $\pm$ 0.3 & 0.57$\pm$0.01& 1.0 $\pm$ 0.2 & 1.2$\pm$0.3 & 0.01 & 0.02\\
1742-078&0.672$\pm$0.008&3.90$\pm$0.05\tablenotemark{b}& 22.7$\pm$0.5 & 8.1 $\pm$ 0.3 & 0.58$\pm$0.01 & 1.3 $\pm$ 0.3 & 1.1$\pm$0.4 & 0.05 & 0.04\\
2148+0657&0.062$\pm$0.002&0.6$\pm$0.1\tablenotemark{a}& 4.3$\pm$0.1 & 0.8 $\pm$ 0.5 & 0.7$\pm$0.1 & $<0.3$ & 3 $\pm$ 2 & 0.006 & 0.008\\
2232+1143&0.062$\pm$0.002&0.6$\pm$0.1\tablenotemark{a}& 4.8$\pm$0.1 & 0.53 $\pm$ 0.51 & 0.8$\pm$0.1 & $<1.0$ & 1.3 $\pm$ 1.2 & 0.002 & 0.0008\\
\hline
\enddata
\tablenotetext{a}{Estimated by E(B-V) using $\mathrm{N_H /E(B-V) = (9.4 \pm 1.6) \times 10^{21} cm^{-2} mag^{-1}}$.}
\tablenotetext{b}{Estimated by E(B-V) using $\mathrm{N_H /E(B-V) = 5.8 \times 10^{21} cm^{-2} mag^{-1}}$.}
\tablenotetext{c}{Opacity corrected.}
\end{deluxetable*}

\section{The CO X-factor}\label{sec:x-factor}
The integrated intensity of CO emission ($\mathrm{W_{CO}}$) is considered to be a good tracer of H$_2$ column density. The CO-to-H$_2$ conversion factor ($\mathrm{X_{CO}}$), which is defined as $\mathrm{X_{CO}}$ = $\mathrm{N_{H_2}}$/$\mathrm{W_{CO }}$, is important for understanding the evolution of star formation, and especially for external galaxies. Due to the distance of extra-galactic objects, CO may be the primary or the only molecular tracer available to explore star formation and determine the total gas mass.

\begin{figure}
\includegraphics[width=1.0\linewidth]{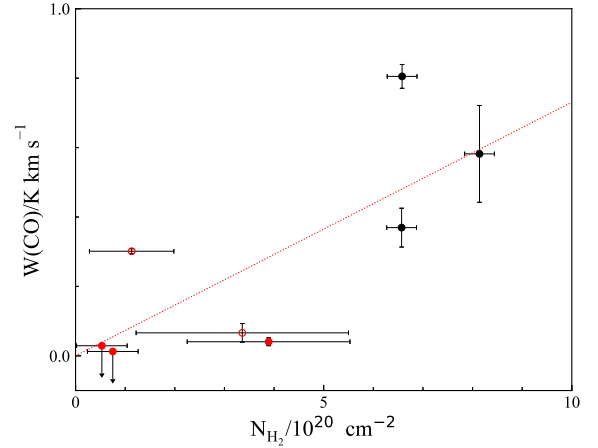}
\caption{The relation of the H$_2$ column density and the integrated intensity of CO (1--0). Red and black points denote sightlines using $\mathrm{N_H /E(B-V) }$ conversion factor of \citealt{Nguyen2018} and \citealt{Bohlin1978}, respectively. Empty red circles denote sightlines with \hi\ opacity correction. The red dotted line shows the fit results corresponding to X-factor (points with upper limit are not included).}
\label{fig:xfactor}
\end{figure}

We have reconstructed the CO integrated intensity from the $\tau$ and $\mathrm{T_{ex}}$ obtained from our RADEX modeling.
The relation between $\mathrm{W_{CO}}$ and $\mathrm{N_{H_2}}$ is shown in Figure \ref{fig:xfactor}.
The average X-factor we obtain in our sightlines is (14 $\pm$ 3) $\times$ 10$^{20}$ cm$^{-2}$ (K\,\kms)$^{-1}$, which is 6 times larger than the average value in the Milky Way \citep{Bolatto2013}. Our results are in agreement with observations in regions without $\mathrm{^{12}CO}$ and $\mathrm{^{13}CO}$ emission detections in Taurus \citep[$\rm{1.2 \times 10^{21} cm^{-2}}$ (K\,\kms)$^{-1}$ in ][]{Pineda2010}. Numerical simulations by \citealt{Shetty2011b} presented $\mathrm{X_{CO}}$-$\mathrm{N_{H_2}}$ relations in molecular clouds with different conditions (e.g., density, metallicity), and they found a high X-factor ($\sim 10^{21}$ cm$^{-2}$ (K \kms)$^{-1}$) where CO abundance is low (10$^{-6}$). In addition, numerical simulations and observations also suggest that turbulence, visual extinction, and metallicity will all influence the CO X-factor \citep{Bell2006a,Shetty2011b}. 
This result indicates that the X-factor should be handled carefully in different physical environments, otherwise the H$_2$ column density may be underestimated in diffuse gas.

\section{Conclusion}\label{sec:conclusion}
We have performed ALMA observations toward 13 continuum sources. The unprecedented sensitivity provided by ALMA allows us to probe cold molecular gas down to a $\mathrm{\tau_{RMS}^{CO}}$ of 0.002 and a $\mathrm{\tau_{RMS}^{HCO^+}}$ of 0.0008. Absorption in both the CO (1--0) and \hcop (1--0) lines were detected toward 6 of 13 sources. In addition, \hcop\ absorption was detected toward two bandpass calibrators (2148+0657 and 2232+1143). Two higher transitions of CO (2--1) and \hcop (3--2) were observed and detected towards 3C454.3.
The main conclusions are as follows:

1. Twelve CO (1--0) components and twenty-three \hcop (1--0) components are detected, revealing CO in diffuse gas with extinction as low as $\mathrm{A_v}\sim$ 0.32\,mag, where no CO emission was detected.

2. By modeling multiple absorption lines -- CO (1--0) and (2--1), and \hcop\ (1--0) and (3--2) -- toward 3C454.3, we derived excitation temperatures of 4.1\,K for CO and 2.7\,K for \hcop. Those temperatures are comparable to the continuum background temperature, making them extremely hard to detect in emission surveys. 

3. The derived column densities of CO and \hcop\ show a positive correlation, while the abundance ratio [CO]/[\hcop] increases by an order of magnitude as the CO column density goes from lower to higher. $\mathrm{N_{HCO^+}}$ has tight linear relation with $\mathrm{N_{H_2}}$ while $\mathrm{N_{CO}}$ does not. Two sightlines without CO detection are found to be rich in \hcop. All facts are consistent with the understanding that, in this intermediate extinction region, \hcop\ is a better tracer of total molecular gas than CO.

4. Toward these absorption sight-lines, we derived the CO X-factor to be (14 $\pm$ 3) $\times$ 10$^{20}$ cm$^{-2}$ (K \kms)$^{-1}$, 6 times larger than the typical value found in the Milky Way. This result is consistent with simulations of clouds going through HI-H$_2$ transitions, where the CO is still building up abundance and has not yet locked up most of the carbon in the ISM.

\acknowledgments
This work has been supported by the National Natural Science Foundation of China No.\ 11988101, No.\ 11725313, No.\ 11803051, National Key R\&D Program of China No.\ 2017YFA0402600, the CAS Strategic Priority Research Program No.\ XDB23000000, and the CAS International Partnership Program No.\ 114A11KYSB20160008.
N.Y. Tang acknowledges the support from the CAS "Light of West China" Program and Young Researcher Grant of National Astronomical Observatories, Chinese Academy of Sciences.
J. R. Dawson acknowledges the support of an Australian Research Council (ARC) DECRA Fellowship (project number DE170101086). L. Bronfman and Ricardo Finger acknowledge support from CONICYT project Basal AFB-170002.

This paper makes use of the following ALMA data: ADS/JAO.ALMA$\#$2015.1.00503.S. ALMA is a partnership of ESO (representing its member states), NSF (USA) and NINS (Japan), together with NRC (Canada) and NSC and ASIAA (Taiwan) and KASI (Republic of Korea), in cooperation with the Republic of Chile. The Joint ALMA Observatory is operated by ESO, AUI/NRAO and NAOJ.

\vspace{5mm}

\software{CASA \citep{MuMullin2007},
         RADEX \citep{Van2007},
         }

\bibliography{reference}

\end{sloppypar}
\end{document}